\documentclass[twocolumn,trackchanges]{aastex701}

\usepackage{graphicx}	
\usepackage{amsmath}	
\usepackage{amssymb}	
\usepackage{xcolor}
\usepackage{color}
\usepackage{soul}
\usepackage{tabularx}

\usepackage{subcaption}
\usepackage{comment}
\usepackage{array}
\usepackage{makecell}
\usepackage{xcolor}

\begin{document}

\title[3X J0042 \textit{NuSTAR} pulsation detection]{Detection of hard X-ray pulsations in non-ULX X-ray pulsar 3X J0042 in M31}

\author[orcid=0009-0005-9417-7193]{Ketan Rikame}
\affiliation{Astronomy and Astrophysics, Raman Research Institute, C.V. Raman Avenue, Bangalore 560080 Karnataka, India}
\email[show]{ketan.rikame@rri.res.in}

\author[orcid=0000-0002-8775-5945]{Biswajit Paul}
\affiliation{Astronomy and Astrophysics, Raman Research Institute, C.V. Raman Avenue, Bangalore 560080 Karnataka, India}
\email{bpaul@rri.res.in}

\begin{abstract}
We report the detection of hard X-ray pulsations from the M31 accreting non-ULX X-ray pulsar 3X J0042 (3XMM J004232.1+411314) using archival \textit{NuSTAR} observations. The source has a spin period of $\sim$3~s and an orbital period of $\sim$4.15~hr, with coherent pulsations previously detected only in \textit{XMM-Newton} observations. We reanalyse all five publicly available \textit{NuSTAR} observations by correcting the photon arrival times for binary motion and optimizing the orbital epoch and projected semi-major axis. Coherent pulsations are detected in all five observations. The broad-band pulse profiles show similar morphology across the five observations spanning $\sim$2.3~yr, with no significant long-term evolution. The pulse profile also shows energy dependence, with structured, non-sinusoidal profiles below $\sim$20~keV and a less structured profile at higher energies. The pulse periods show an overall decrease over the $\sim$2.3~yr \textit{NuSTAR} baseline, continuing the long-term spin-up trend observed in the $\sim$16~yr of \textit{XMM-Newton} observations, but with a faster average spin-up rate of $\sim$$-1.1\times10^{-10}$~s\,s$^{-1}$ compared to the earlier reported $\sim$$-6\times10^{-11}$~s\,s$^{-1}$. The short orbital period and inferred low-mass companion star shows it to be a low-mass X-ray binary containing a magnetized neutron star, with systems such as Her X-1, 3A 1822-371, and XMM J174457-2850.3 providing closer Galactic comparisons. This work represents the first detection of hard X-ray pulsations from a non-ULX extragalactic accreting pulsar.

\end{abstract}





\section{Introduction}

Accreting X-ray pulsars with high magnetic fields generally exhibit hard X-ray spectra extending to tens of keV. Their pulse profiles and pulsed fractions can show significant energy dependence, providing information about the geometry of the accretion column and the physical processes governing the emission \citep{Paul2002,Becker2005,Pradhan2021}. Extending studies of accretion-powered pulsars to extragalactic systems is particularly valuable, as it provides an opportunity to investigate whether their timing and spectral properties are similar to those of the well-studied Galactic population. Detection of pulsations across the broad X-ray band allows the energy dependence of the pulse properties to be studied, providing a direct comparison with Galactic high-magnetic-field accreting pulsars.

Prior to this work, hard X-ray pulsations from extragalactic accreting neutron stars had been detected primarily in ultraluminous X-ray sources (ULXs). The discovery of pulsations from M82 X-2 with \textit{NuSTAR} provided direct evidence that a ULX could be powered by an accreting neutron star \citep{Bachetti2014}, followed by the identification of additional pulsating ULXs such as NGC 7793 P13 and NGC 5907 ULX1 \citep{Furst2016,Israel2017}. These discoveries demonstrate the capability of hard X-ray observations to reveal pulsations from accreting neutron stars beyond the Milky Way. They also suggest that counterparts of Galactic hard X-ray pulsars in nearby galaxies may be detectable in the hard X-ray band, provided that their flux and pulsation amplitudes are sufficiently high.

M31 provides a nearby environment in which such studies can be extended. Two accreting X-ray pulsar systems, 3XMM J004232.1+411314 (hereafter 3X J0042) and 3XMM J004301.4+413017 (hereafter 3X J0043), had previously been identified through pulsations in \textit{XMM-Newton} observations \citep{Esposito2016, Rodr_guez_Castillo_2018}. The $\sim$1.2 s pulsar 3X J0043 has an orbital period of $\sim$1.27 d, while 3X J0042 has a $\sim$3 s spin period and a much shorter $\sim$4.15 hr orbit. Although pulsations in these systems are detectable in the soft-to-medium X-ray band with \textit{XMM-Newton}, their hard X-ray pulsation properties are not known. In this context, the hard X-ray brightness of 3X J0042, which dominates the M31 bulge emission above $\sim$25 keV, makes it a particularly suitable target for such a search.

3X J0042 is an X-ray binary located in the bulge of M31. The source has been detected in several \textit{Chandra}, \textit{XMM-Newton}, \textit{Swift}, and \textit{NuSTAR} observations and reaches X-ray luminosities of a few $10^{38}$~erg~s$^{-1}$, making it one of the most luminous persistent X-ray sources in M31 \citep{Marelli2017}. Using simultaneous \textit{NuSTAR} and \textit{Swift} observations, \citet{Yukita_2017} identified 3X J0042 as the hard X-ray counterpart of the source Swift J0042.6+4112 and showed that it dominates the hard X-ray emission from the M31 bulge above $\sim$25~keV. The broadband X-ray spectrum was found to consist of a soft thermal component together with a hard power-law continuum and a high-energy cutoff, consistent with the spectral properties of accreting X-ray pulsars with high magnetic fields \citep{Yukita_2017}. Despite its strong hard X-ray emission, however, no significant pulsations were detected in the available \textit{NuSTAR} observations.

A systematic analysis of archival \textit{XMM-Newton} observations by \citet{Marelli2017} revealed recurrent dipping episodes in the X-ray light curve, recurring on a timescale of $\sim$4~hr. These dips were interpreted as orbital modulation in a high-inclination low-mass X-ray binary, providing evidence for a compact binary system and suggesting an orbital period of $\sim$4~hr. \citet{Rodr_guez_Castillo_2018} subsequently detected coherent $\sim$3~s pulsations from 3X J0042 in \textit{XMM-Newton} observations and, through pulse-arrival-time analysis, determined an orbital period of $\sim$4.15~hr, consistent with the periodic dipping phenomena \citep{Marelli2017}. The barycentric and orbit-corrected pulse period was found to decrease by $\sim$28 ms over a baseline of approximately 16 yr, corresponding to an average spin-up rate of $\dot{P}\sim-6\times10^{-11}$ s\,s$^{-1}$. In addition to this long-term spin-up, shorter-timescale variations in the pulse period were observed and were suggested to be associated with changes in the X-ray luminosity \citep{Rodr_guez_Castillo_2018}.

Despite the discovery of the pulsar and the determination of its orbital parameters, coherent pulsations have not been reported in the \textit{NuSTAR} observations that originally identified 3X J0042 as the dominant hard X-ray source in M31. For a compact binary with an orbital period of only $\sim$4.15~hr, orbital smearing can significantly reduce the detectability of coherent pulsations if photon arrival times are not corrected for the binary motion. The availability of the orbital solution derived by \citet{Rodr_guez_Castillo_2018} therefore motivates a re-examination of the \textit{NuSTAR} data. In this work, we reanalyse all available \textit{NuSTAR} observations of 3X J0042.

The paper is structured as follows. Section~\ref{sec:analysis} describes the \textit{NuSTAR} observations, data reduction procedures, and timing analysis. Sections \ref{sec:orbital_corr} and \ref{sec:orbital_opt} describe orbital motion correction and its optimization. The subsequent coherent pulsation search is described in Section~\ref{sec:pulsation_opt}, followed by the energy-resolved pulse-profile analysis in Section~\ref{sec:energy_res}. Section~\ref{discussion} discusses the hard X-ray pulsation detection in 3X J0042, the energy dependence and stability of the pulse profiles, the nature of the system, and its spin evolution. Section~\ref{Conclusion} provides a summary and conclusions.

\section{Observations, Data Reduction, and analysis}
\label{sec:analysis}

\begin{figure*}
    \centering
    \begin{subfigure}{0.38\textwidth}
        \centering
        \includegraphics[width=\linewidth]{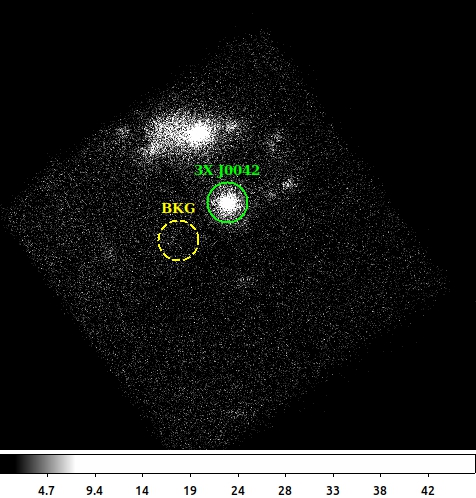}
        \label{fig:fig1}
    \end{subfigure}
    \hfill
    \begin{subfigure}{0.58\textwidth}
        \centering
        \includegraphics[width=\linewidth]{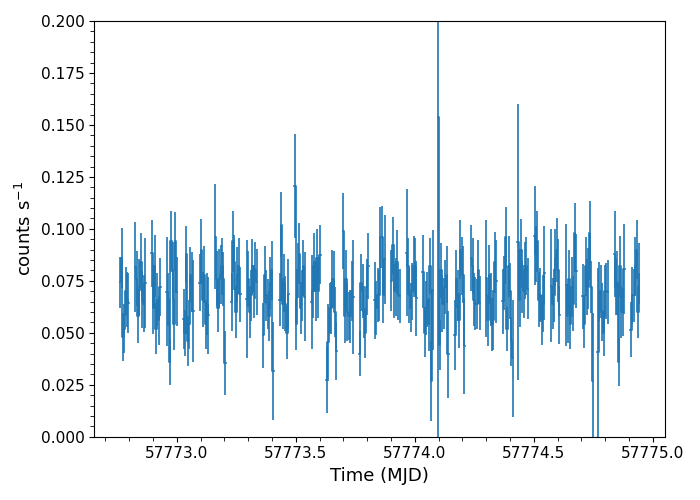}
        \label{fig:fig2}
    \end{subfigure}

\caption{\textit{NuSTAR} observation (OBSID 50201001002) of 3X J0042. The left panel shows the \textit{NuSTAR} image, with the source extraction region of radius 50~arcsec indicated by a green circle and the nearby background extraction region of the same radius indicated by a yellow dashed circle. The right panel shows the corresponding background-subtracted 5-20 keV light curve, binned at 500~s.}

\label{fig:50201001002_img_lc}
\end{figure*}

There are five archival \textit{NuSTAR} observations of M31 that include 3X J0042, as listed in Table~\ref{tab:obs_list}. We have analyzed all five observations in this work. The data were processed using the \texttt{nupipeline} script available in HEASoft version 6.35.1 with CALDB version 20220331. Source events were extracted from a circular region with a radius of 50~arcsec centered on the source position, while background events were extracted from a nearby source-free circular region of the same radius. Barycentric corrections were applied to the event data using the FTOOL \texttt{barycorr}. Figure~\ref{fig:50201001002_img_lc} shows a representative \textit{NuSTAR} image with the source extraction region, along with the corresponding barycenter-corrected light curve for the longest observation (OBSID: 50201001002).

The spectral properties of 3X J0042 have been extensively studied using previous \textit{NuSTAR} and \textit{Swift} observations by \citet{Yukita_2017}. Since the focus of the present work is the timing properties and hard X-ray pulsations of the source, we do not perform or report an independent spectral analysis here and adopt the previously reported spectral results for context.

For timing studies of accreting X-ray pulsars, pulse arrival times are commonly measured from a sequence of relatively short data segments and subsequently modelled by accounting for the delays introduced by the orbital motion and intrinsic changes in the pulse period. In the present case, however, the relatively low count rate of 3X J0042 in the \textit{NuSTAR} observations, together with its short orbital period of $\sim$4.15~hr, limits the sensitivity of such an approach when applied to short data segments. We therefore adopt a strategy in which the orbital correction parameters are optimized directly using the source event data over the entire observation, followed by a coherent pulsation search using the full available observation. This approach maximizes the photon statistics available for the pulsation search while allowing the orbital parameters affecting the timing correction to be optimized directly from the \textit{NuSTAR} data.

\begin{table}
\centering
\caption{\textit{NuSTAR} observations of 3X J0042 analyzed in this work.}
\label{tab:obs_list}
\fontsize{7}{7}\selectfont
\setlength{\tabcolsep}{2.5pt}
\begin{tabular}{
>{\centering\arraybackslash}p{1.4cm}
>{\centering\arraybackslash}p{1.7cm}
>{\centering\arraybackslash}p{0.8cm}
>{\centering\arraybackslash}p{1.0cm}
>{\centering\arraybackslash}p{0.8cm}
>{\centering\arraybackslash}p{1.0cm}
}
\hline
OBSID & Observation Start & Duration & On-source Time & 
\multicolumn{2}{>{\centering\arraybackslash}p{1.8cm}}{Average 5-20 keV count rate} \\
      & (yyyy-mm-dd hh:mm:ss) & (ks) & (ks) &
\multicolumn{2}{c}{($10^{-2}$ counts s$^{-1}$)} \\
\cline{5-6}
      & & & & Source & Background \\
\hline
50101001002 & 2015-09-13 16:43:09 & 186.2 & 105.4 & 6.22 & 0.81 \\
50201001002 & 2017-01-19 18:13:20 & 188.9 & 110.4 & 7.92 & 0.88 \\
50302001002 & 2017-06-28 10:49:55 & 78.2  & 45.2  & 5.00 & 0.86 \\
50302001004 & 2017-07-07 10:43:34 & 81.6  & 44.3  & 5.26 & 0.78 \\
50302001006 & 2018-01-04 03:33:23 & 78.7  & 45.9  & 4.52 & 0.47 \\
\hline
\end{tabular}
\end{table}

\subsection{Orbital Motion Correction}
\label{sec:orbital_corr}

The orbital motion of the neutron star introduces orbital phase dependent delays in the pulse arrival times, which can significantly reduce the detectability of coherent pulsations in a compact binary system if left uncorrected. We therefore corrected the event arrival times for binary motion using the orbital timing solution reported by \citet{Rodr_guez_Castillo_2018} before performing the pulsation search. The binary correction primarily depends on following geometric parameters: the orbital epoch, the projected semi-major axis ($a_{\rm x}\sin i$), and the orbital period ($P_{\rm orb}$). Although these parameters were previously determined from \textit{XMM-Newton} observations by \citet{Rodr_guez_Castillo_2018}, not all of them can be directly adopted for the \textit{NuSTAR} observations due to relatively large uncertainties associated with the parameter estimation. The projected semi-major axis reported by \citet{Rodr_guez_Castillo_2018}, $a_{\rm x}\sin i = 0.59\pm0.04$ lt-s, has a relatively large uncertainty, which can influence the accuracy of the orbital correction. The orbital period was measured to be 4.15 hours with an uncertainty of approximately 1\%, while the epoch of the ascending node is measured with an uncertainty of around 260~s \citep{Rodr_guez_Castillo_2018}. These uncertainties accumulate over the large number of orbital cycles separating the \textit{XMM-Newton} and \textit{NuSTAR} observations. Consequently, the orbital epoch cannot be propagated reliably to the epochs of the \textit{NuSTAR} observations. However, each individual \textit{NuSTAR} observation spans only a few orbital cycles and therefore provides limited sensitivity to the orbital period itself. We therefore fix the orbital period to the value reported by \citet{Rodr_guez_Castillo_2018} throughout this analysis, while optimizing the orbital epoch and the projected semi-major axis to improve the orbital motion correction and, consequently, the detectability of coherent pulsations, as described below.

\subsection{Orbital Parameter Optimization}
\label{sec:orbital_opt}

\begin{figure*}
    \centering
    \begin{subfigure}{0.49\textwidth}
        \centering
        \includegraphics[width=\linewidth]{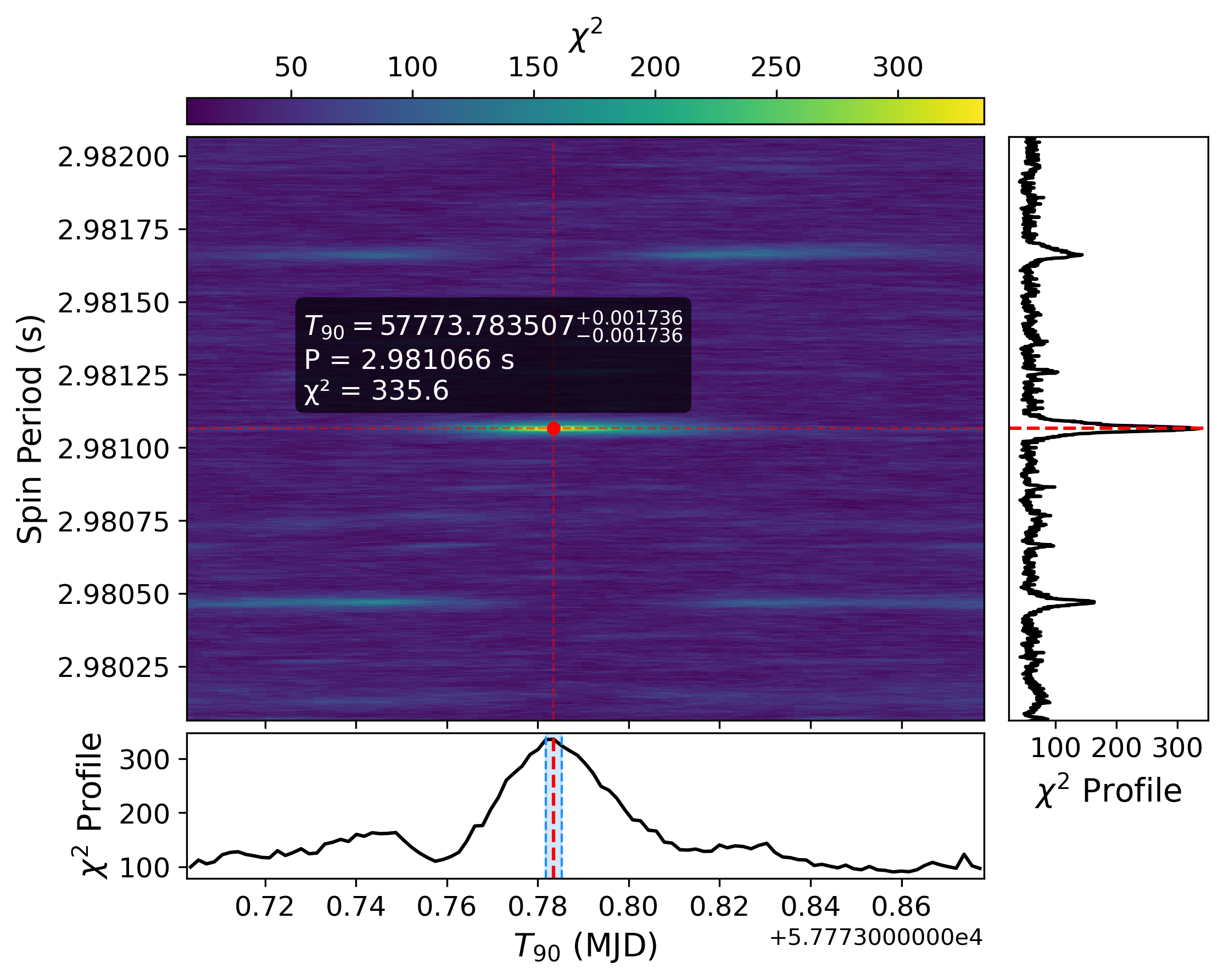}
        \label{fig:fig1}
    \end{subfigure}
    \hfill
    \begin{subfigure}{0.49\textwidth}
        \centering
        \includegraphics[width=\linewidth]{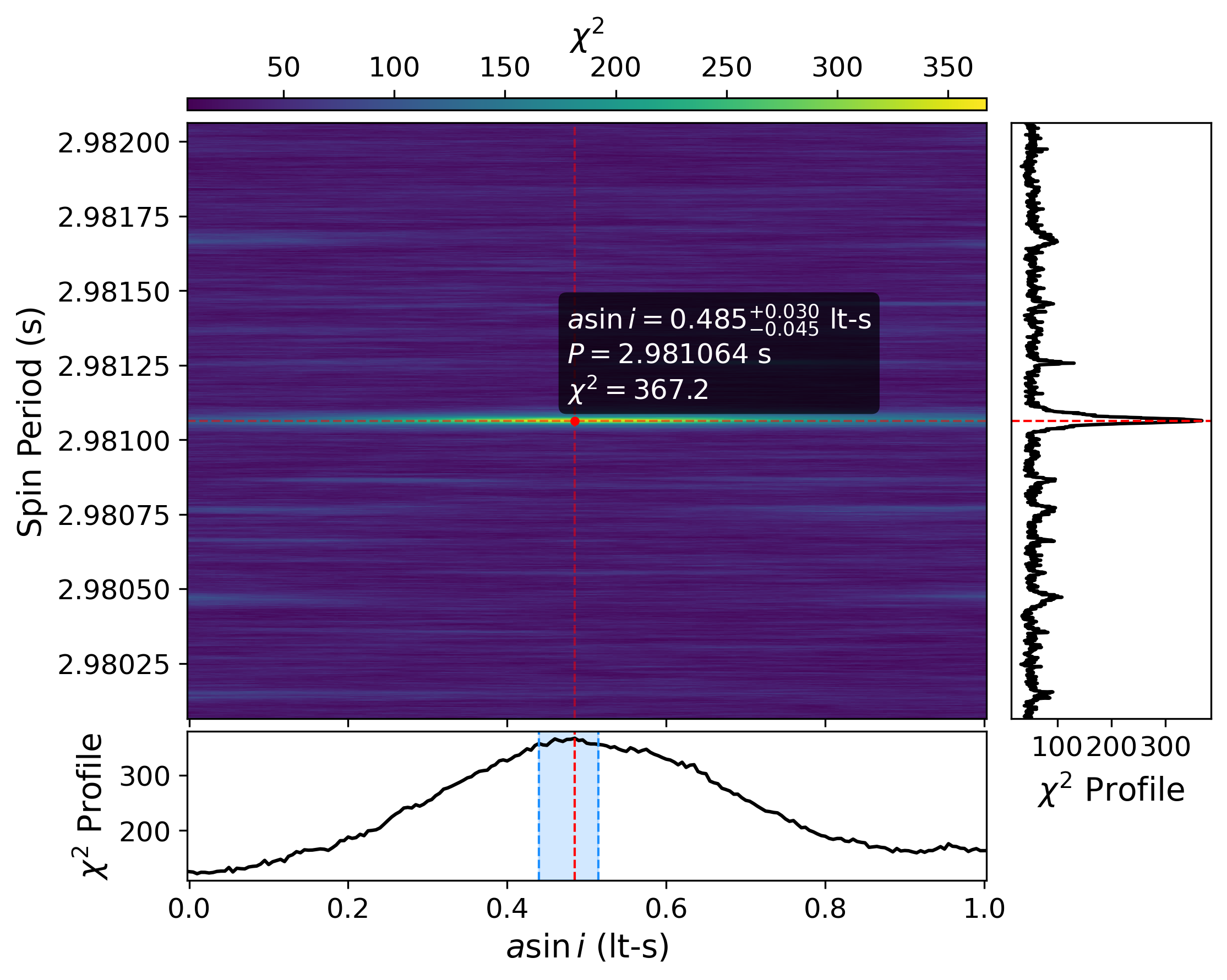}
        \label{fig:fig2}
    \end{subfigure}

    \caption{
Initial orbital-parameter optimization procedure for the \textit{NuSTAR} observation (OBSID 50201001002) of 3X J0042. The initial search over $T_{90}$ was performed using $a_{\rm x}\sin i=0.59$~lt-s \citep{Rodr_guez_Castillo_2018}. The resulting optimized $T_{90}$ was then used to optimize $a_{\rm x}\sin i$. The corresponding maximum epoch-folding statistic ($\chi^2$) is shown as a function of the trial $T_{90}$ and $a_{\rm x}\sin i$ in the bottom subpanels of the left and right panels, respectively.
}
    \label{fig:initial_optimization}
\end{figure*}

The initial optimization of the orbital parameters was carried out using the two longest \textit{NuSTAR} observations (OBSID 50101001002 and 50201001002), which provide the largest number of source photons. Given the short orbital period of 3X J0042, we assume a circular orbit for the binary timing correction. The orbital delay in a circular binary orbit can be described as,
\begin{equation}
\Delta t_{\rm orb} =
(a_{\rm x}\sin i)
\cos\left[
2\pi\left(
\frac{t-T_{90}}{P_{\rm orb}}
\right)
\right],
\label{eq:orbital_delay}
\end{equation}
where $a_{\rm x}\sin i$ is the projected semi-major axis expressed in light-seconds, $P_{\rm orb}$ is the orbital period, and $T_{90}$ is the orbital epoch corresponding to an orbital longitude of $90^{\circ}$ and to the maximum orbital delay \citep{Jain2024}. The binary-corrected photon arrival time is then given by
\begin{equation}
t_{\rm corr}=t-\Delta t_{\rm orb}.
\label{eq:binary_corrected_time}
\end{equation}

The orbital correction is strongly dependent on the orbital phase, and hence on the value of $T_{90}$, particularly when extrapolated over a large number of orbital cycles. As an initial step, a grid search over the orbital epoch ($T_{90}$) was performed by correcting the event arrival times for trial values of $T_{90}$ spanning one orbital cycle, while keeping the other orbital parameters fixed at the values reported by \citet{Rodr_guez_Castillo_2018}. For each trial value of $T_{90}$, an epoch-folding search was carried out using the HEASoft tool \texttt{efsearch}, and the value of $T_{90}$ corresponding to the largest $\chi^{2}$ was adopted as the best-fit orbital epoch for that observation. Using the optimized orbital epoch, we subsequently optimized the projected semi-major axis in a similar manner. The photon arrival times were corrected for a grid of trial values of $a_{\rm x}\sin i$, while keeping the optimized value of $T_{90}$ fixed, and the pulsation search was repeated for each trial value.

Figure~\ref{fig:initial_optimization} shows the initial orbital-parameter optimization procedure for OBSID 50201001002. The initial $T_{90}$ search was performed using the published value of $a_{\rm x}\sin i$, followed by an optimization of $a_{\rm x}\sin i$ using the optimized value of $T_{90}$. Repeating the optimization procedure did not produce any further increase in the maximum $\chi^2$, nor did it result in significant changes in the optimized values of $T_{90}$ and $a_{\rm x}\sin i$. The same procedure was independently applied to the other long \textit{NuSTAR} observation, OBSID 50101001002. Both \textit{NuSTAR} observations yielded consistent measurements of the projected semi-major axis, with $a_{\rm x}\sin i = 0.490^{+0.060}_{-0.045}$~lt-s for OBSID 50101001002 and $0.485^{+0.030}_{-0.045}$~lt-s for OBSID 50201001002. The quoted uncertainties are at the $3\sigma$ confidence level. These values are consistent with each other and are both smaller than the value of $a_{\rm x}\sin i = 0.59\pm0.04$~lt-s (at $1\sigma$ confidence) reported by \citet{Rodr_guez_Castillo_2018}. We therefore adopt the weighted mean of the two \textit{NuSTAR} measurements for the analysis of the remaining observations.

After optimizing the projected semi-major axis, the orbital epoch was determined independently for all five \textit{NuSTAR} observations following the procedure described above. For each observation, the photon arrival times were corrected using a grid of trial values of $T_{90}$, while keeping the projected semi-major axis fixed at the optimized value and the orbital period fixed at the value reported by \citet{Rodr_guez_Castillo_2018} (See Figure \ref{fig:timing_workflow}, Panel (a)).

\subsection{Pulsation Detection Optimization}
\label{sec:pulsation_opt}

\begin{figure*}
    \centering

    \begin{subfigure}{0.49\textwidth}
        \centering
        \includegraphics[width=\linewidth]{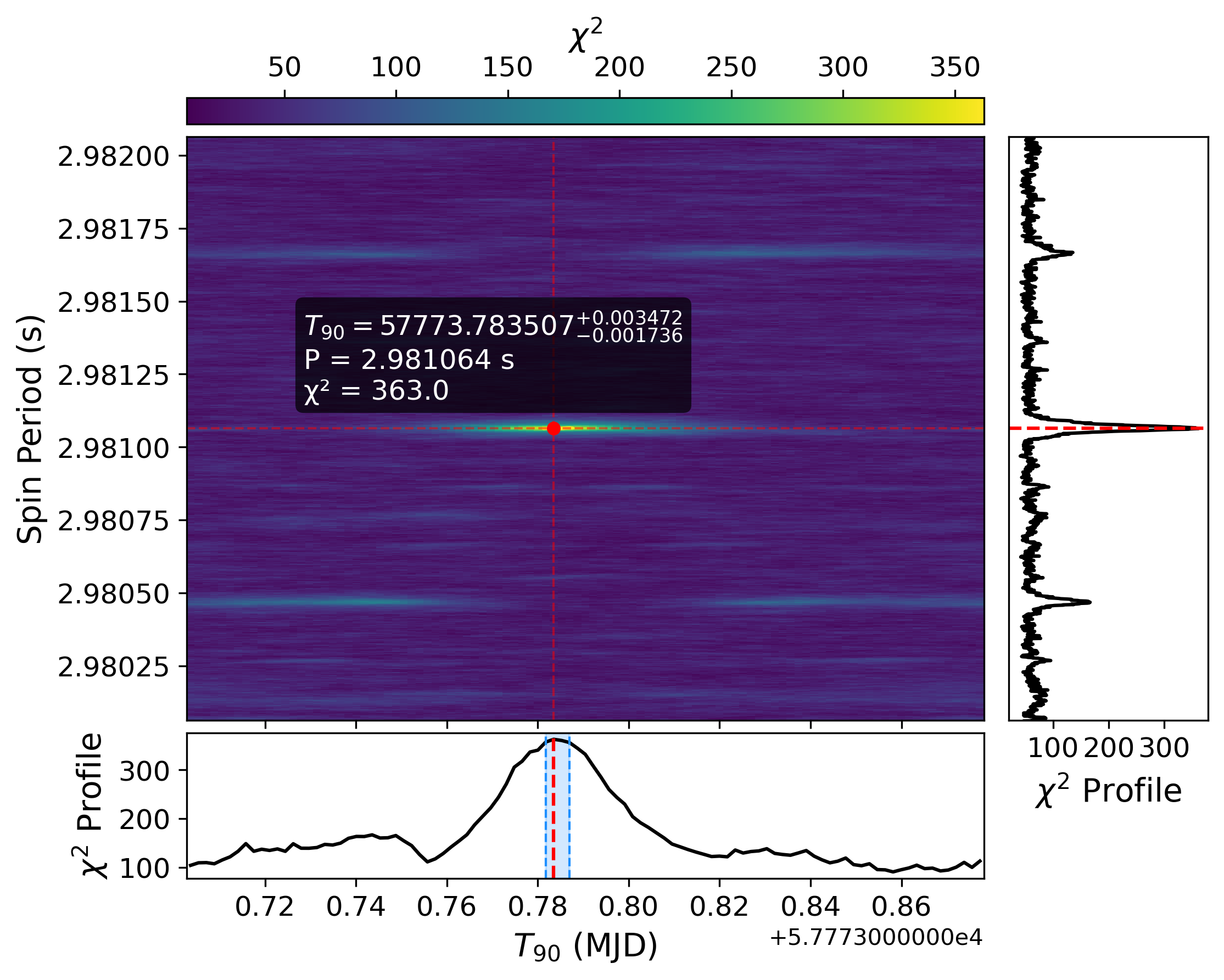}
        \caption{}
        \label{fig:t90_scan}
    \end{subfigure}
    \hfill
    \begin{subfigure}{0.49\textwidth}
        \centering
        \includegraphics[width=\linewidth]{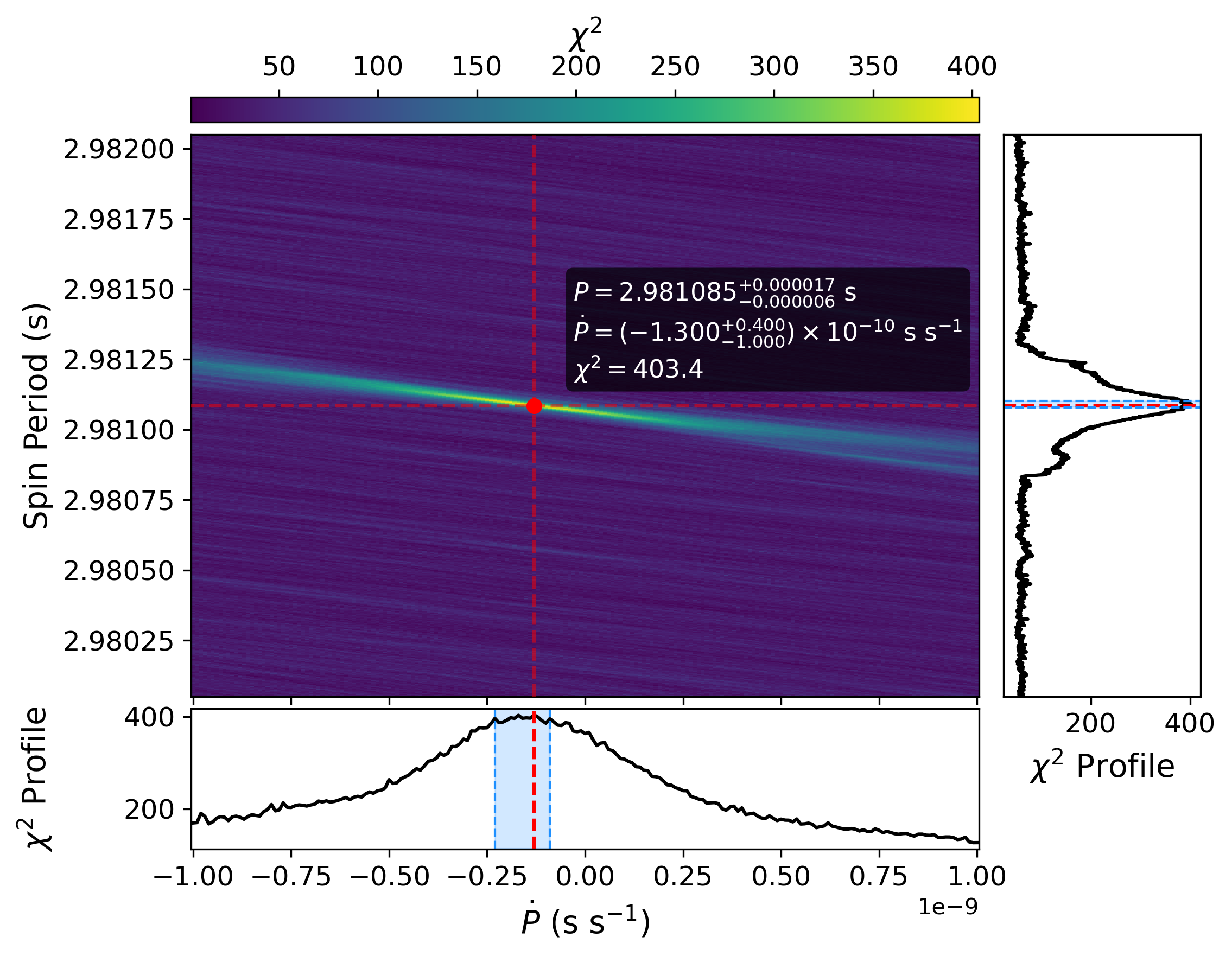}
        \caption{}
        \label{fig:pdot_search}
    \end{subfigure}

    \vspace{0.5cm}

    \begin{subfigure}{0.49\textwidth}
        \centering
        \includegraphics[width=\linewidth]{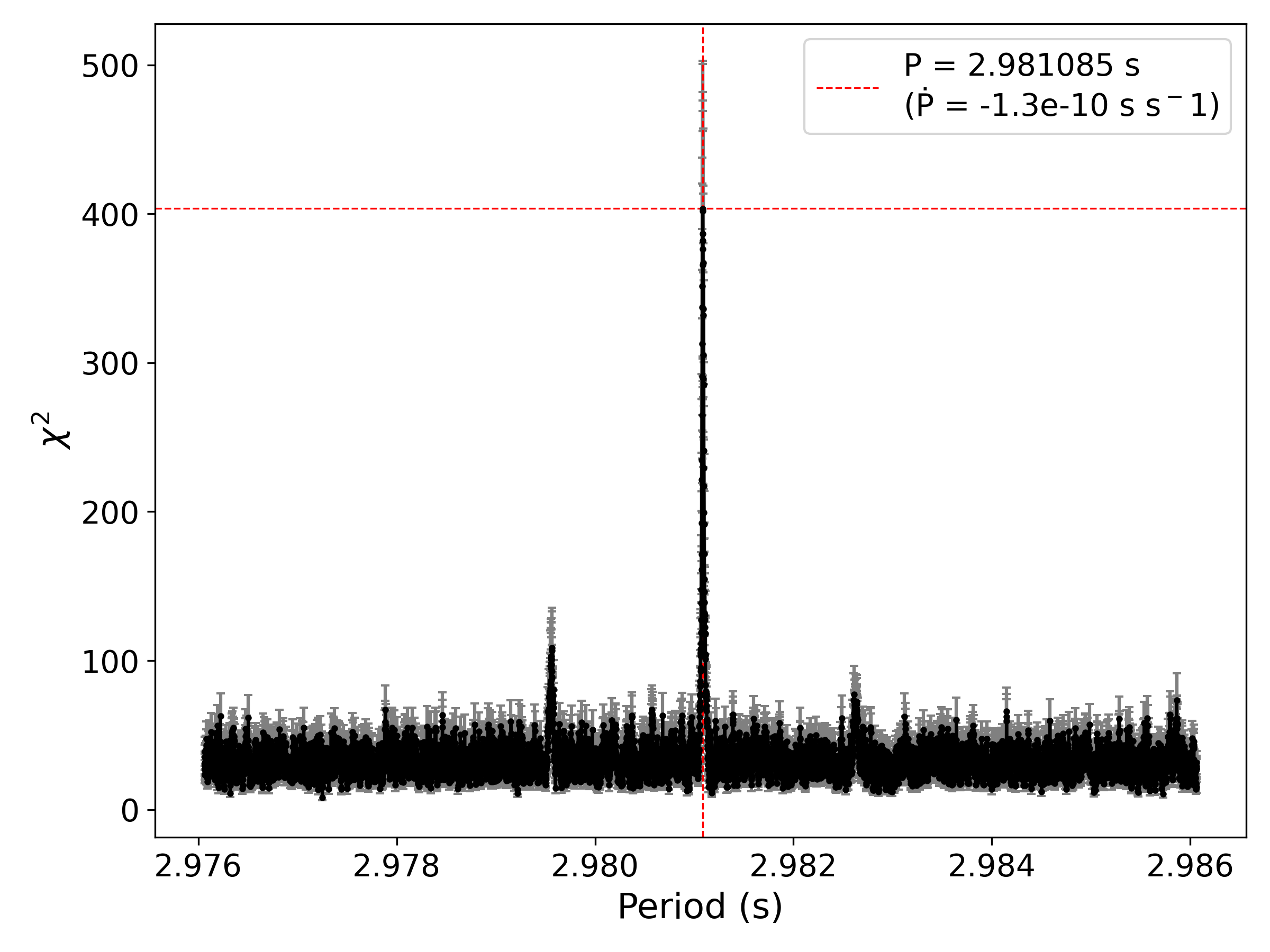}
        \caption{}
        \label{fig:efsearch}
    \end{subfigure}
    \hfill
    \begin{subfigure}{0.49\textwidth}
        \centering
        \includegraphics[width=\linewidth]{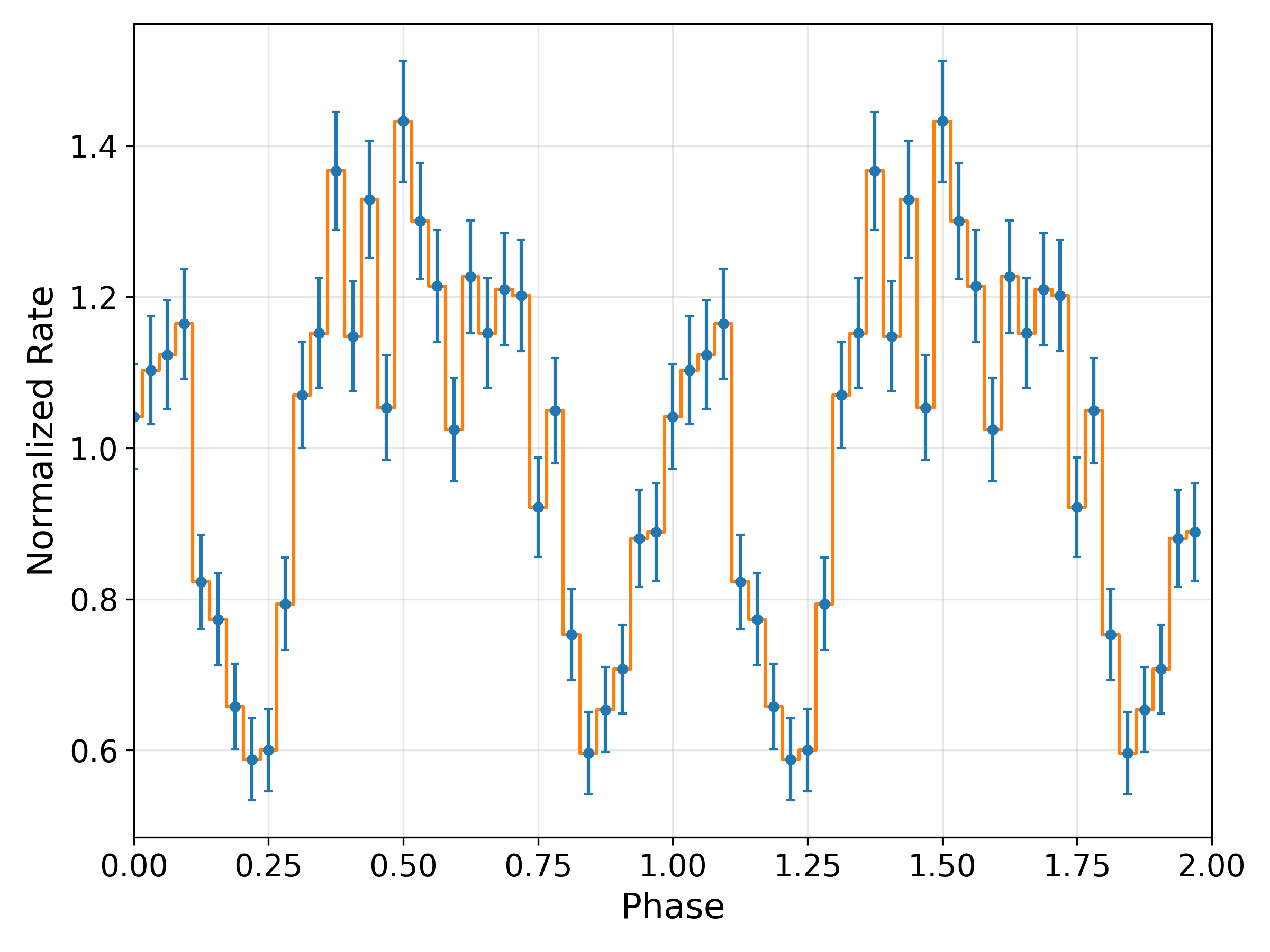}
        \caption{}
        \label{fig:profile}
    \end{subfigure}

    \caption{
Timing analysis of the \textit{NuSTAR} observation (OBSID: 50201001002) of 3X J0042. 
(\textbf{a}) Epoch-folding ($\chi^2$) search after applying orbital corrections for trial values of $T_{90}$ spanning one orbital cycle, with the projected semi-major axis fixed at the optimized value and the orbital period fixed at the value reported by \citet{Rodr_guez_Castillo_2018}. The strongest pulsation signal is obtained for $T_{90}=57774.6516$ MJD. 
(\textbf{b}) Two-dimensional search over pulse period and trial period derivative ($\dot{P}$) in the range $-1\times10^{-9}$ to $1\times10^{-9}$ s\,s$^{-1}$. The optimal solution corresponds to $P=2.981085s$ and $\dot{P}=-1.3\times10^{-10}$ s\,s$^{-1}$. 
(\textbf{c}) Epoch-folding search for the optimized orbital parameters. (\textbf{d}) Corresponding normalized pulse profile.
    }
    \label{fig:timing_workflow}
\end{figure*}

Once the optimal orbital correction had been established, a refined search for the neutron-star spin parameters was carried out. The pulse period was first determined using the HEASoft task \texttt{efsearch} over a narrow range around the expected spin period. The period corresponding to the maximum epoch-folding statistic was adopted as the best estimate of the pulse period. The \textit{XMM-Newton} observations showed both a long-term spin-up of the neutron star, with an average period derivative of $\dot{P}\sim-6\times10^{-11}$~s\,s$^{-1}$ over a baseline of around 16~yr, as well as shorter-timescale variations in the pulse period \citep{Rodr_guez_Castillo_2018}. Therefore, we also searched for a pulse-period derivative in each individual \textit{NuSTAR} observation. A two-dimensional search over pulse period and period derivative was performed around this best-fit period by exploring trial values of $\dot{P}$ in the range $-1\times10^{-9}$ to $1\times10^{-9}$ s\,s$^{-1}$. The combination of $P_{\rm spin}$ and $\dot{P}$ that maximized the epoch-folding statistic was adopted as the final timing solution for each observation. Figure~\ref{fig:timing_workflow} illustrates this timing analysis procedure for OBSID 50201001002, showing the orbital-epoch search, the subsequent $P_{\rm spin}$--$\dot{P}$ search, and the resulting pulse profile. We note that each individual observation covers several binary orbits ($\sim$13 orbits for each of the OBSIDs 50101001002 and 50201001002, and $\sim$5 orbits for each of the remaining three observations), and therefore, the orbital parameter estimates presented in Section~\ref{sec:orbital_opt} are unlikely to be significantly affected by an underlying pulse-period derivative.

The timing optimization was performed using the source event data. After determining the optimal orbital and spin parameters, the source and background events were folded using the corresponding timing solution to obtain the pulse profiles. The folded background profiles were found to be relatively flat, with no additional features that could contribute to the source pulse profiles. We therefore treated the background contribution as a constant count rate and subtracted the corresponding background level from the source pulse profiles to obtain the final background-subtracted pulse profiles.

\begin{table*}
\centering
\setlength{\tabcolsep}{6pt}
\renewcommand{\arraystretch}{1.5}
\caption{Timing results obtained from the five \textit{NuSTAR} observations of 3X J0042.}
\label{tab:timing_results}
\fontsize{8}{8}\selectfont
\begin{tabular}{>{\centering\arraybackslash}p{2.5cm}>{\centering\arraybackslash}p{3.5cm}>{\centering\arraybackslash}p{3.0cm}>{\centering\arraybackslash}p{2.0cm}>{\centering\arraybackslash}p{2.0cm}>{\centering\arraybackslash}p{2.0cm}}
\hline
OBSID
&
\shortstack[c]{$T_{90}$\\(MJD; errors in seconds)}
&
$P_{\rm spin}$ (s)
&
\shortstack[c]{$\dot{P}$\\($10^{-10}$ s\,s$^{-1}$)}
&
\shortstack[c]{Pulse Fraction$^{*}$\\(\%)}
&
Maximum $\chi^2$
\\
\hline
50101001002 & $57279.61939^{+100\,\mathrm{s}}_{-300\,\mathrm{s}}$ & $2.9855384^{+0.0000109}_{-0.0000084}$ & $-0.60^{+0.60}_{-0.70}$ & 34.0 & 225.7 \\
50201001002 & $57773.78351^{+300\,\mathrm{s}}_{-150\,\mathrm{s}}$ & $2.9810850^{+0.0000170}_{-0.0000060}$ & $-1.30^{+0.40}_{-1.00}$ & 37.2 & 403.4 \\
50302001002 & $57932.84427^{+400\,\mathrm{s}}_{-100\,\mathrm{s}}$ & $2.9794052^{+0.0000580}_{-0.0000154}$ & $0.85^{+2.05}_{-6.65}$ & 49.0 & 143.9 \\
50302001004 & $57941.77112^{+400\,\mathrm{s}}_{-350\,\mathrm{s}}$ & $2.9793161^{+0.0000084}_{-0.0000235}$ & $-3.00^{+2.45}_{-1.15}$ & 52.6 & 174.5 \\
50302001006 & $58122.46788^{+350\,\mathrm{s}}_{-300\,\mathrm{s}}$ & $2.9773511^{+0.0000158}_{-0.0000116}$ & $4.30^{+2.10}_{-3.95}$ & 44.9 & 127.9 \\
\hline
\multicolumn{6}{l}{\footnotesize
$^{*}$ Pulse fraction is defined as the ratio of the difference between the maximum and minimum count rate in a pulse profile to their sum.
}
\end{tabular}
\end{table*}

\begin{figure}
    \centering
    \includegraphics[width=\columnwidth]{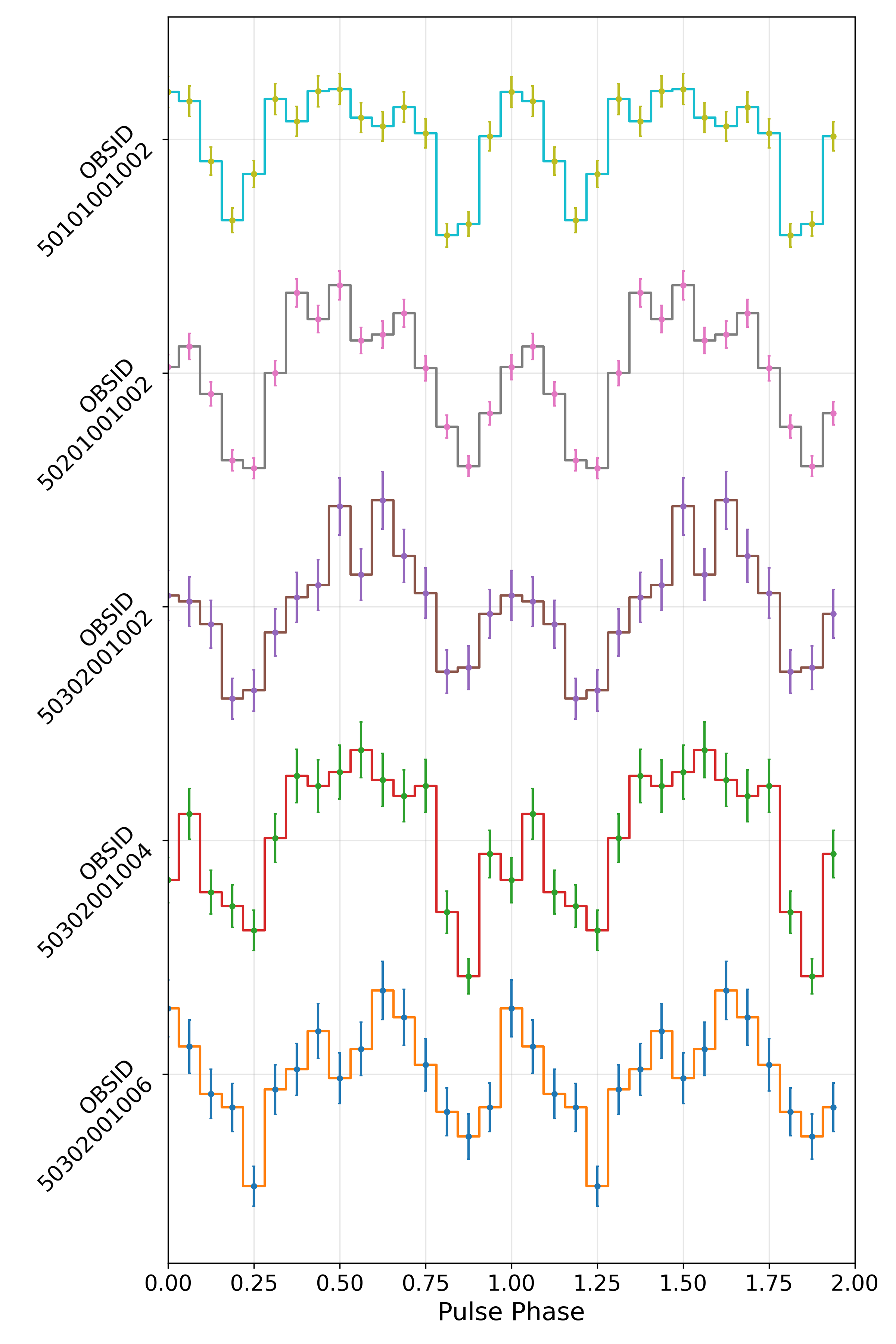}
    \caption{
    Phase-aligned pulse profiles obtained from the five \textit{NuSTAR} observations of 3X J0042 after applying the optimized orbital and spin parameters. The profiles are shown with arbitrary vertical offsets for visual clarity. Coherent pulsations are seen in all five observations.
    }
    \label{fig:aligned_profiles}
\end{figure}

\begin{figure}
    \centering
    \includegraphics[width=\columnwidth]{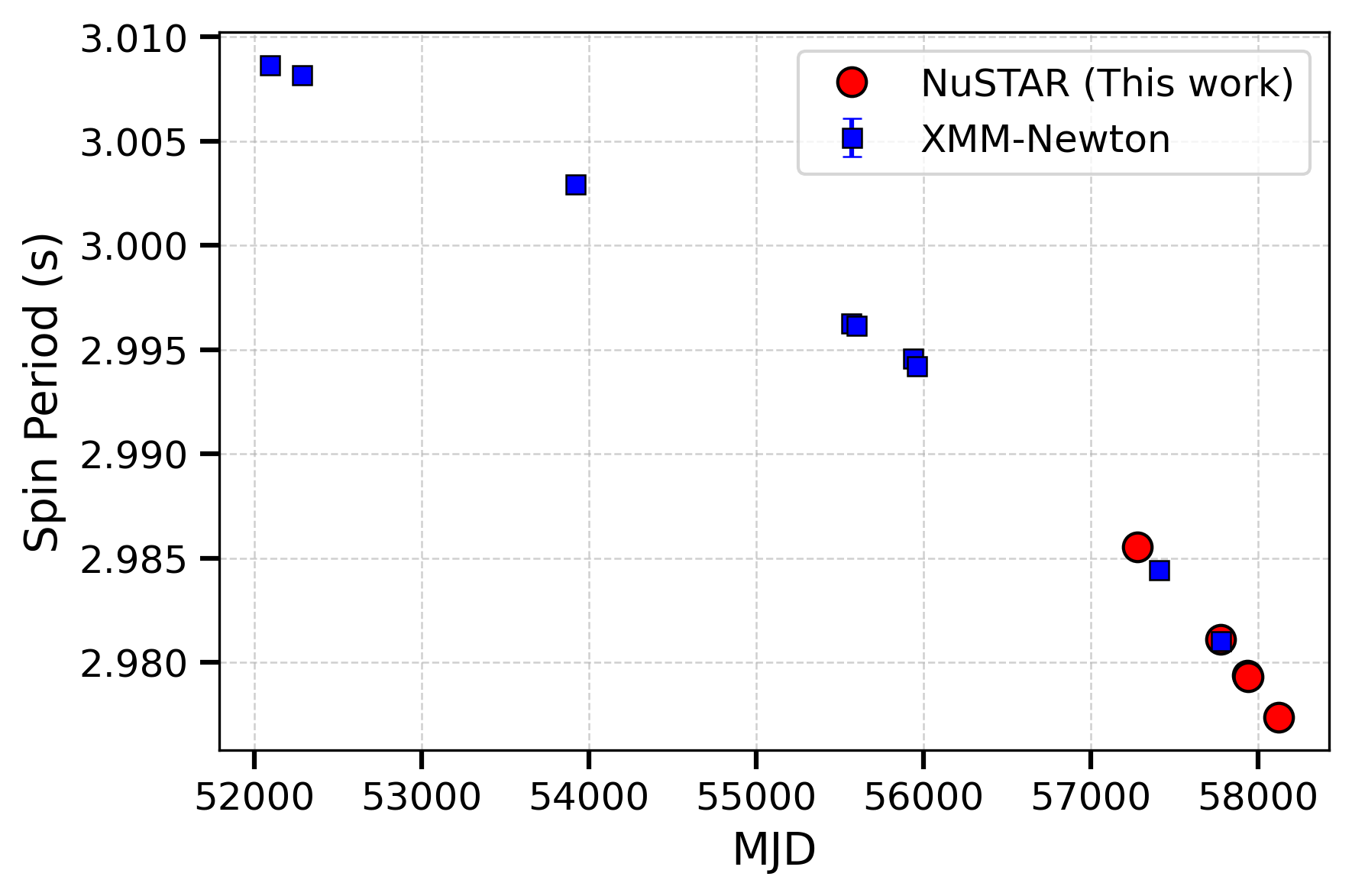}
    \caption{
    Long-term evolution of the spin period of 3X J0042. The pulse periods measured from the five \textit{NuSTAR} observations are shown by red circles, while the earlier \textit{XMM-Newton} measurements are shown by blue squares. The \textit{NuSTAR} measurements continue the long-term spin-up trend established from the \textit{XMM-Newton} observations.
    }
    \label{fig:spin_evolution}
\end{figure}

The optimized orbital epochs, spin-period periods, spin-period derivatives, and corresponding maximum epoch-folding statistics obtained from the five \textit{NuSTAR} observations are listed in Table~\ref{tab:timing_results}. Coherent pulsations are detected in all five observations after applying the optimized orbital correction, as demonstrated by the phase-aligned pulse profiles shown in Figure~\ref{fig:aligned_profiles}. It can be seen that overall pulse morphology remain similar across observations spanning around 2.3~yr. The pulse periods show an overall decrease with time, indicating continued spin-up of the neutron star, as shown in Figure~\ref{fig:spin_evolution}.

\subsection{Energy Resolved Profiles}
\label{sec:energy_res}

\begin{figure}
\centering

\begin{subfigure}[t]{0.5\linewidth}
    \centering
    \includegraphics[width=\linewidth]{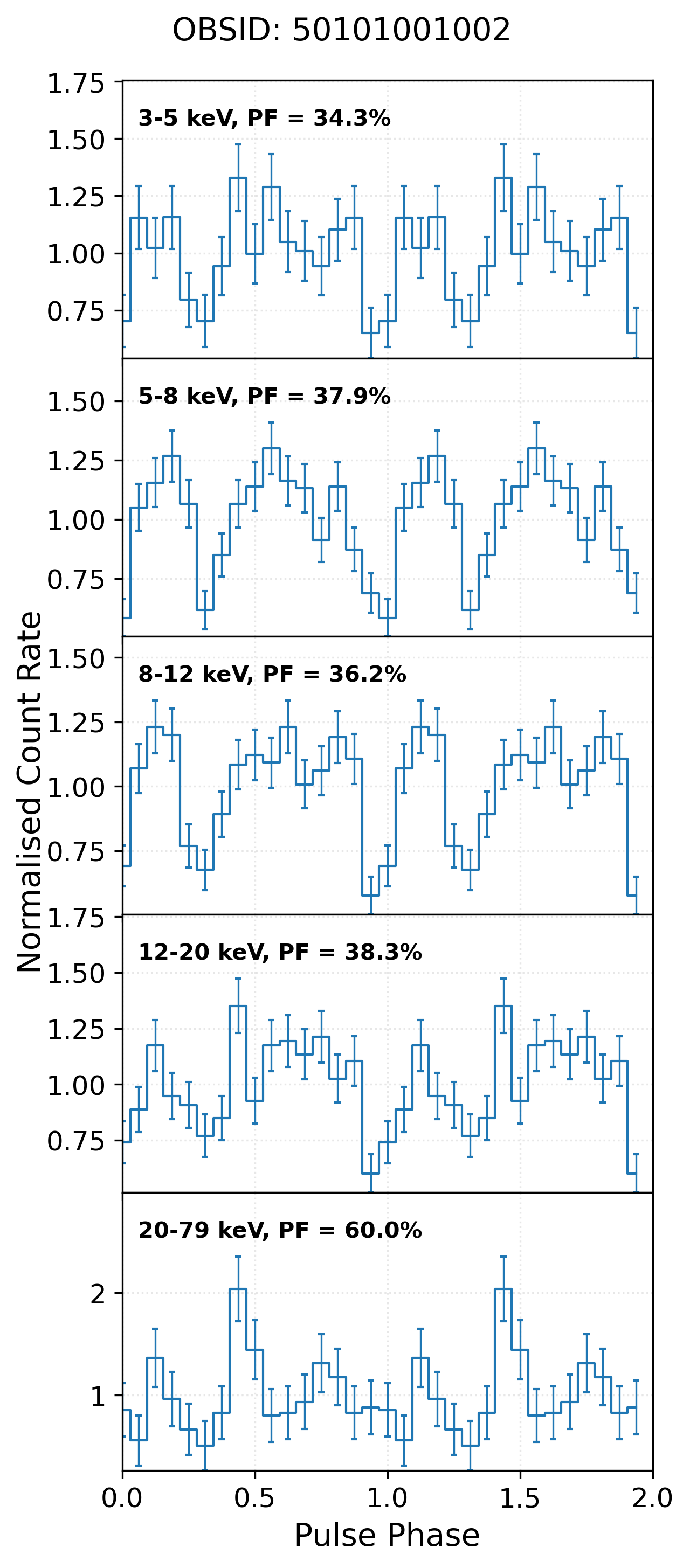}
\end{subfigure}\hfill
\begin{subfigure}[t]{0.5\linewidth}
    \centering
    \includegraphics[width=\linewidth]{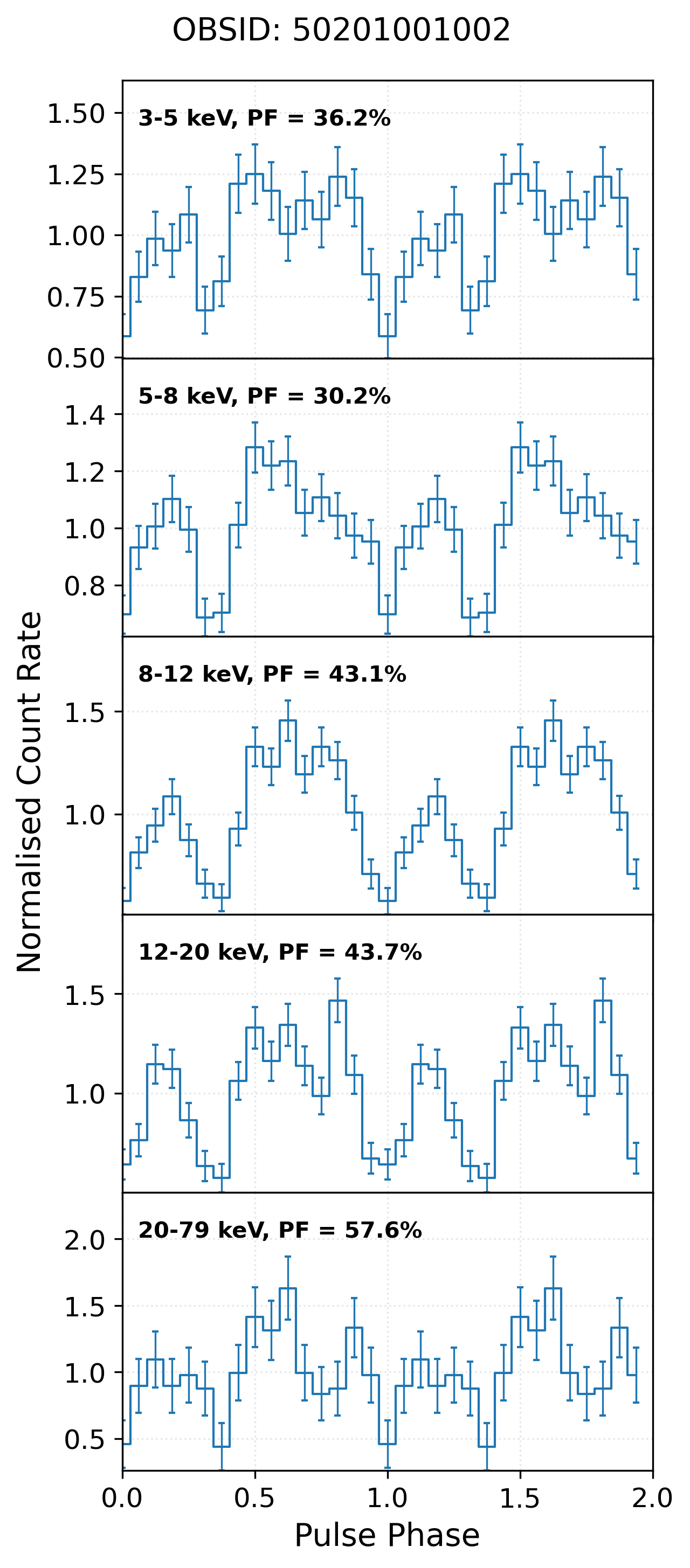}
\end{subfigure}

\caption{Energy-resolved pulse profiles of 3X J0042 for the two longest \textit{NuSTAR} observations, OBSID 50101001002 (left) and 50201001002 (right). From top to bottom, the profiles correspond to the 3-5, 5-8, 8-12, 12-20, and 20-79~keV energy bands. The pulse profiles show a structured, non-sinusoidal morphology below $\sim$20~keV, while the profiles above 20~keV appear less structured owing to the lower photon statistics.
}
\label{fig:energy_profiles}
\end{figure}

To investigate the energy dependence of the pulsation morphology, we generated pulse profiles in five energy bands: 3-5, 5-8, 8-12, 12-20, and 20-79~keV. The energy-resolved profiles were generated for the two longest \textit{NuSTAR} observations, OBSID 50101001002 and 50201001002, and are shown in Figure~\ref{fig:energy_profiles}. The pulse profiles in each observation are shown vertically for the different energy bands.

The pulse profiles show a structured and non-sinusoidal morphology at energies below $\sim$20~keV. In particular, multiple peaks and minima are visible in the 5-8, 8-12, and 12-20~keV bands. The overall morphology is broadly similar between the two observations in the corresponding energy bands, although the relative amplitudes of individual features show some variation. At higher energies (20-79~keV), the pulse profiles appear less structured, with the individual features becoming less pronounced. This could be related to changes in the relative contributions of the different emission components with energy, although the limited photon statistics at higher energies may also contribute to the observed reduction in structure.

\begin{figure}
    \centering
    \includegraphics[width=\columnwidth]{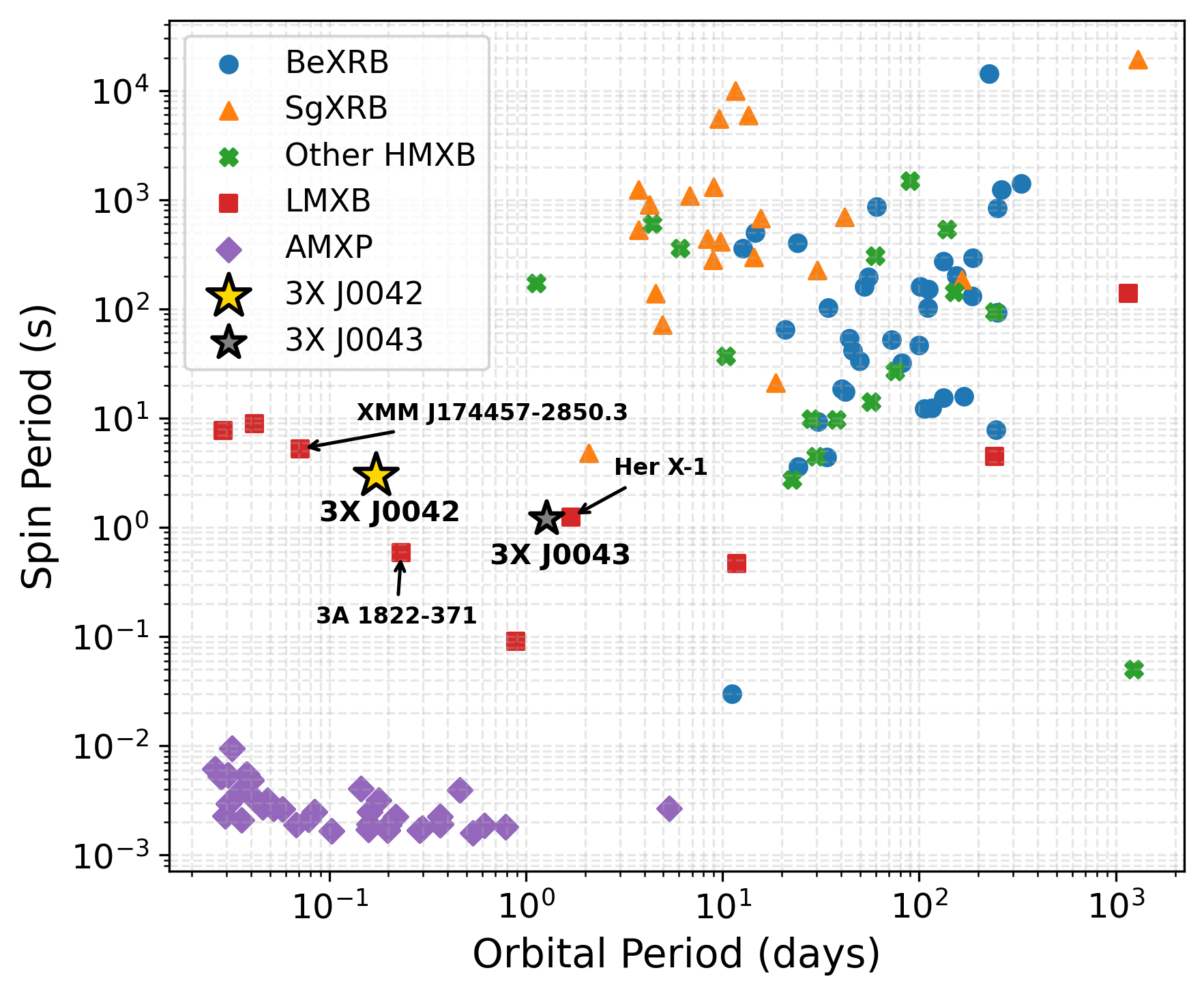}
    \caption{Spin-orbital period (Corbet) diagram showing the distribution of Galactic accreting X-ray pulsars. The positions of the M31 pulsars 3X J0042 and 3X J0043 are marked by yellow and gray stars, respectively. Galactic accreting pulsars located close to these M31 systems in the diagram are also annotated for comparison.}
    \label{fig:corbet_diagram}
\end{figure}

\section{Discussion}
\label{discussion}

We have reanalyzed data from all available \textit{NuSTAR} observations of 3X J0042 by accounting for the orbital motion of the neutron star. By optimizing the orbital epoch and the projected semi-major axis, we recover coherent pulsations in all five \textit{NuSTAR} observations. This represents the first detection of coherent hard X-ray pulsations from 3X J0042 and directly connects the source dominating the hard X-ray emission from the M31 bulge with the accreting neutron-star pulsar previously identified through \textit{XMM-Newton} timing observations. The detection also establishes 3X J0042 as the first non-ULX extragalactic accreting pulsar detected through coherent hard X-ray pulsations.

\subsection{Hard X-ray pulsations from an extragalactic accreting pulsar}

The detection of hard X-ray pulsations from 3X J0042 extends the population of extragalactic accreting hard X-ray pulsars beyond the pulsating ultraluminous X-ray sources (ULXs). Pulsations from several extragalactic ULXs have been detected with \textit{NuSTAR}, demonstrating that accreting neutron stars can be identified through their hard X-ray pulsations at extragalactic distances. However, 3X J0042 is a persistent, non-ULX accreting X-ray pulsar in M31. The detection is therefore particularly relevant for establishing whether the hard X-ray properties of Galactic accreting pulsars are also observed in neutron-star binaries in nearby galaxies. The hard X-ray detection is also important in the context of the spectral properties of 3X J0042. The source has a broadband spectrum characterized by a soft thermal component together with a hard continuum extending to tens of keV, with a high-energy cutoff, which is broadly similar to the spectra observed from Galactic accreting X-ray pulsars. The detection of pulsations in the same hard X-ray band demonstrates that the hard component is directly associated with the X-ray pulsar.

\subsection{Temporal Stability of the Pulse Profile}

The phase-aligned pulse profiles from all five \textit{NuSTAR} observations show broadly similar overall morphology despite the observations spanning approximately 2.3~yr (Figure~\ref{fig:aligned_profiles}). Although variations in the relative amplitudes of individual features are present, no pronounced long-term changes in the pulse shape are evident. The persistence of the pulse-profile morphology over this timescale suggests that the geometry of the hard X-ray emitting region remains broadly stable, despite the observed long-term spin-up trend (see Section~\ref{spinup_discussion}).

\subsection{Energy dependence of the pulse profile}

The energy-resolved pulse profiles show change in the pulse morphology across the \textit{NuSTAR} energy band. Below $\sim$20~keV, the profiles show a structured and distinctly non-sinusoidal morphology, with multiple peaks and minima. On the other hand, the profiles in the 20-79~keV band appear less structured, although the larger statistical uncertainties at high energies make the morphology more difficult to constrain. The reduction in the apparent structure at high energies may therefore result from limited photon statistics, changes in the relative contributions of different emission components, or a combination of both effects. Up to $\sim$20~keV, the pulsed fraction values remain comparable, consistent with the \textit{XMM-Newton} observations, which showed no significant energy dependence of the pulsed fraction up to 12~keV \citep{Rodr_guez_Castillo_2018}.

\subsection{Orbital configuration and comparison with Galactic X-ray pulsars}

The Corbet diagram, which relates the spin and orbital periods of accreting X-ray pulsars, provides a useful empirical framework for comparing different classes of X-ray binaries \citep{Corbet1984,Corbet1986}. Galactic accreting X-ray pulsars span a wide range of spin and orbital periods and occupy distinct regions in the spin-orbital period plane depending on the nature of the companion star and the dominant mode of accretion. In Figure~\ref{fig:corbet_diagram}, we show the positions of 3X J0042 \citep[and this work]{Rodr_guez_Castillo_2018} and 3X J0043 \citep{Esposito2016} together with Galactic HMXB and LMXB pulsars from the catalogues of \citet{hmxb_cat} and \citet{lmxb_cat}, respectively.

The spin period of $\sim$3~s and orbital period of $\sim$4.15~hr place 3X J0042 at the position shown by the yellow star in the spin-orbital period (Corbet) diagram (Figure~\ref{fig:corbet_diagram}). Its orbital period is much shorter than those of classical wind-fed and disk-fed high-mass X-ray binaries, which typically have orbital periods of several days to hundreds of days. At the same time, its spin period is longer than those of Galactic accreting millisecond pulsars. The short orbital period and relatively slow spin of 3X J0042 are consistent with a low-mass X-ray binary containing a magnetized neutron star. The periodic dips and short orbital period reported previously also support a compact, high-inclination system \citep{Marelli2017,Rodr_guez_Castillo_2018}. The orbital solution further suggests a low-mass companion of approximately $0.2$-$0.3\,M_{\odot}$ for the assumed neutron-star mass and inclination \citep{Rodr_guez_Castillo_2018}.

Among the Galactic accreting pulsars located near to 3X J0042 in the Corbet diagram, Her X-1 provides a useful comparison because it has a low/intermediate-mass companion and a disk-fed accretion geometry, with a spin period of $\sim$1.24~s and an orbital period of $\sim$1.7~d \citep{Tananbaum1972, Wilson1997}. The other M31 pulsar, 3X J0043, has a spin period of $\sim$1.2~s and an orbital period of $\sim$1.27~d \citep{Esposito2016}, placing it even closer to Her X-1 in the Corbet diagram. The other nearby systems highlighted in Figure~\ref{fig:corbet_diagram}, 3A 1822-371 and XMM J174457-2850.3, also provide useful comparisons with short-period accreting neutron-star binaries. 3A 1822-371 is a dipping and eclipsing low-mass X-ray binary with a spin period of $\sim$0.59~s and an orbital period of $\sim$5.57~hr \citep{Jonker2001,Burderi2010}. XMM J174457-2850.3 is another compact Galactic X-ray binary associated with the Galactic Centre, with an inferred orbital period of around 1-2~hr \citep{Degenaar2014,Heinke2015}.

The location of 3X J0042 in the Corbet diagram, together with its short orbital period, dipping behaviour, and inferred low-mass companion, distinguishes it from the classical high-mass X-ray pulsar population and places it among the short-period accreting neutron-star systems.

\subsection{Long-term spin evolution}
\label{spinup_discussion}

The spin periods measured from the five \textit{NuSTAR} observations show an overall decrease with time, indicating continued long-term spin-up of the neutron star (Figure~\ref{fig:spin_evolution}). Over the $\sim$2.3~yr \textit{NuSTAR} baseline, 3X J0042 shows an average spin-period derivative of $\dot{P} \sim -1.1\times10^{-10}$~s\,s$^{-1}$. An average period derivative of $\dot{P} \sim -6\times10^{-11}$~s\,s$^{-1}$ was reported from the earlier \textit{XMM-Newton} observations over a baseline of around 16~yr \citep{Rodr_guez_Castillo_2018}. The \textit{NuSTAR} measurements therefore extend the previously observed long-term spin-up trend to a later epoch and indicate a faster spin-up rate during the \textit{NuSTAR} epoch.

Although the overall trend is one of spin-up, the individual \textit{NuSTAR} measurements show variations in the inferred period derivative, including both positive and negative values. Such short-timescale variations may reflect changes in the accretion torque, as also suggested by the shorter-timescale pulse-period variations observed with \textit{XMM-Newton}. However, the earlier period measurements were obtained in the softer X-ray band with \textit{XMM-Newton}, whereas the later measurements presented here are from the hard X-ray band with \textit{NuSTAR}. We therefore do not attempt to investigate whether the higher spin-up rate during the \textit{NuSTAR} epoch is associated with an increase in the X-ray luminosity. Continued timing observations with consistent broadband coverage over a longer baseline will be required to investigate the relation between spin-up rate and X-ray luminosity.

\subsection{Orbital ephemeris}

This work provides improved measurements of the orbital ephemeris at the epochs of the individual \textit{NuSTAR} observations. The optimized values of $T_{90}$ allow the orbital correction to be established locally for each observation and are essential for recovering the pulsations. The projected semi-major axis was also optimized using the two longest \textit{NuSTAR} observations. Both observations give consistent values of $a_{\rm x}\sin i$ of around $0.49$~lt-s, which is lower than the value of $0.59\pm0.04$~lt-s reported from the \textit{XMM-Newton} observations by \citet{Rodr_guez_Castillo_2018}. However, these measurements cannot presently be combined into a new orbital ephemeris with sufficient confidence to improve the orbital period. In particular, the uncertainties in the measured orbital epochs do not allow the number of complete orbital cycles between the \textit{NuSTAR} observations to be determined unambiguously. As a result, assigning an absolute cycle count to the individual $T_{90}$ measurements would introduce an ambiguity in the derived orbital period. A longer temporal baseline combining the available \textit{XMM-Newton} and \textit{NuSTAR} observations, together with additional future observations, would be valuable for resolving this ambiguity. Such a dataset could provide a sufficiently precise orbital ephemeris to investigate possible changes in the orbital period and to characterize the long-term orbital dynamics of the system. Improved measurements of both the orbital and spin evolution would also help constrain the accretion geometry and evolutionary state of this unusual extragalactic X-ray pulsar.

In the Milky Way, the known population of accretion-powered X-ray pulsars is dominated by high-mass X-ray binaries, while most accreting millisecond X-ray pulsars are transient systems. In contrast, the two currently known accretion-powered pulsars in M31, 3X J0042 and 3X J0043, are both associated with low-mass companions. This difference may indicate that the population of accretion-powered pulsars in M31 are significantly different from that in the Milky Way.

\section{Conclusion}
\label{Conclusion}

We have analyzed data form all five archival \textit{NuSTAR} observations of the M31 X-ray pulsar 3X J0042 and searched for coherent pulsations after correcting the photon arrival times for binary motion.

\begin{enumerate}
    \item Coherent pulsations with a spin period of around 3s are detected in all five \textit{NuSTAR} observations. This represents the first detection of hard X-ray pulsations from 3X J0042 and provides direct evidence that the source dominating the hard X-ray emission from the M31 bulge is the accreting neutron-star pulsar previously identified through \textit{XMM-Newton} timing observations.

    \item The measured pulse periods show an overall decrease with time, continuing the long-term spin-up trend established from the earlier \textit{XMM-Newton} observations. The period derivatives measured within individual \textit{NuSTAR} observations show shorter-timescale variations, suggesting variations in the accretion torque in addition to the long-term spin-up.

    \item The pulse profiles from all five \textit{NuSTAR} observations show a structured, non-sinusoidal morphology that remains broadly stable over the $\sim$2.3~yr observational baseline. This overall similarity indicates that the pulse morphology does not undergo pronounced long-term evolution over this timescale.

    \item The energy-resolved pulse profiles show more complex structure below $\sim$20~keV, while the individual features become less pronounced at higher energies. This energy dependence may reflect changes in the relative contributions of different emission components, although the limited photon statistics at higher energies prevent a firm interpretation.

    \item The short orbital period, low-mass companion inferred from the orbital solution, and pulsation and spectral properties of 3X J0042 suggest the presence of a low-mass X-ray binary containing a magnetized neutron star. Its nature is more similar to the disk-fed system Her X-1, although 3X J0042 has a shorter orbital period.
\end{enumerate}

The detection of hard X-ray pulsations from 3X J0042 extends the population of extragalactic hard X-ray pulsars beyond the pulsating ULXs and demonstrates that persistent, non-ULX accreting neutron stars in nearby galaxies can also be detected through their hard X-ray pulsations. Further observations combining \textit{XMM-Newton} and \textit{NuSTAR} timing data over a longer baseline will be important for improving the orbital ephemeris, constraining the long-term spin evolution, and investigating the relation between accretion torque and X-ray luminosity.


\section*{Acknowledgements}
This research has made use of archival data and software provided NASA's High Energy (HEASARC), which is a service of the Astrophysics Science Division at NASA/GSFC and \textit{NuSTAR} data analysis pipeline.

\section*{Data Availability}

The observational data underlying this work is publicly available through the High Energy Astrophysics Science Archive Research Center (HEASARC). Any additional information will be shared on reasonable request to the corresponding author.


\bibliography{main}{}

@article{Rodr_guez_Castillo_2018,
   title={Discovery of a 3 s Spinning Neutron Star in a 4.15 hr Orbit in the Brightest Hard X-Ray Source in M31},
   volume={861},
   ISSN={2041-8213},
   url={http://dx.doi.org/10.3847/2041-8213/aacf40},
   DOI={10.3847/2041-8213/aacf40},
   number={2},
   journal={The Astrophysical Journal Letters},
   publisher={American Astronomical Society},
   author={Rodríguez Castillo, Guillermo A. and Israel, Gian Luca and Esposito, Paolo and Papitto, Alessandro and Stella, Luigi and Tiengo, Andrea and Luca, Andrea De and Marelli, Martino},
   year={2018},
   month=July, pages={L26} }

@article{Yukita_2017,
   title={Identification of the Hard X-Ray Source Dominating the E &gt; 25 keV Emission of the Nearby Galaxy M31},
   volume={838},
   ISSN={1538-4357},
   url={http://dx.doi.org/10.3847/1538-4357/aa62a3},
   DOI={10.3847/1538-4357/aa62a3},
   number={1},
   journal={The Astrophysical Journal},
   publisher={American Astronomical Society},
   author={Yukita, M. and Ptak, A. and Hornschemeier, A. E. and Wik, D. and Maccarone, T. J. and Pottschmidt, K. and Zezas, A. and Antoniou, V. and Ballhausen, R. and Lehmer, B. D. and Lien, A. and Williams, B. and Baganoff, F. and Boyd, P. T. and Enoto, T. and Kennea, J. and Page, K. L. and Choi, Y.},
   year={2017},
   month=Mar, pages={47} }

@ARTICLE{Marelli2017,
       author = {{Marelli}, Martino and {Tiengo}, Andrea and {De Luca}, Andrea and {Salvetti}, David and {Saronni}, Luca and {Sidoli}, Lara and {Paizis}, Adamantia and {Salvaterra}, Ruben and {Belfiore}, Andrea and {Israel}, Gianluca and {Haberl}, Frank and {D'Agostino}, Daniele},
        title = "{Discovery of Periodic Dips in the Brightest Hard X-Ray Source of M31 with EXTraS}",
      journal = {\apjl},
         year = 2017,
        month = dec,
       volume = {851},
       number = {2},
          eid = {L27},
        pages = {L27},
          doi = {10.3847/2041-8213/aa9b2e},
archivePrefix = {arXiv},
       eprint = {1711.05540},
 primaryClass = {astro-ph.HE},
       adsurl = {https://ui.adsabs.harvard.edu/abs/2017ApJ...851L..27M}
}

@ARTICLE{Bachetti2014,
       author = {{Bachetti}, M. and {Harrison}, F.~A. and {Walton}, D.~J. and {Grefenstette}, B.~W. and {Chakrabarty}, D. and {F{\"u}rst}, F. and {Barret}, D. and {Beloborodov}, A. and {Boggs}, S.~E. and {Christensen}, F.~E. and {Craig}, W.~W. and {Fabian}, A.~C. and {Hailey}, C.~J. and {Hornschemeier}, A. and {Kaspi}, V. and {Kulkarni}, S.~R. and {Maccarone}, T. and {Miller}, J.~M. and {Rana}, V. and {Stern}, D. and {Tendulkar}, S.~P. and {Tomsick}, J. and {Webb}, N.~A. and {Zhang}, W.~W.},
        title = "{An ultraluminous X-ray source powered by an accreting neutron star}",
      journal = {\nat},
         year = 2014,
        month = oct,
       volume = {514},
       number = {7521},
        pages = {202-204},
          doi = {10.1038/nature13791},
archivePrefix = {arXiv},
       eprint = {1410.3590},
 primaryClass = {astro-ph.HE},
       adsurl = {https://ui.adsabs.harvard.edu/abs/2014Natur.514..202B}
}

@ARTICLE{Furst2016,
       author = {{F{\"u}rst}, F. and {Walton}, D.~J. and {Harrison}, F.~A. and {Stern}, D. and {Barret}, D. and {Brightman}, M. and {Fabian}, A.~C. and {Grefenstette}, B. and {Madsen}, K.~K. and {Middleton}, M.~J. and {Miller}, J.~M. and {Pottschmidt}, K. and {Ptak}, A. and {Rana}, V. and {Webb}, N.},
        title = "{Discovery of Coherent Pulsations from the Ultraluminous X-Ray Source NGC 7793 P13}",
      journal = {\apjl},
         year = 2016,
        month = nov,
       volume = {831},
       number = {2},
          eid = {L14},
        pages = {L14},
          doi = {10.3847/2041-8205/831/2/L14},
archivePrefix = {arXiv},
       eprint = {1609.07129},
 primaryClass = {astro-ph.HE},
       adsurl = {https://ui.adsabs.harvard.edu/abs/2016ApJ...831L..14F}
}

@ARTICLE{Israel2017,
       author = {{Israel}, Gian Luca and {Belfiore}, Andrea and {Stella}, Luigi and {Esposito}, Paolo and {Casella}, Piergiorgio and {De Luca}, Andrea and {Marelli}, Martino and {Papitto}, Alessandro and {Perri}, Matteo and {Puccetti}, Simonetta and {Castillo}, Guillermo A. Rodr{\'\i}guez and {Salvetti}, David and {Tiengo}, Andrea and {Zampieri}, Luca and {D'Agostino}, Daniele and {Greiner}, Jochen and {Haberl}, Frank and {Novara}, Giovanni and {Salvaterra}, Ruben and {Turolla}, Roberto and {Watson}, Mike and {Wilms}, Joern and {Wolter}, Anna},
        title = "{An accreting pulsar with extreme properties drives an ultraluminous x-ray source in NGC 5907}",
      journal = {Science},
         year = 2017,
        month = feb,
       volume = {355},
       number = {6327},
        pages = {817-819},
          doi = {10.1126/science.aai8635},
archivePrefix = {arXiv},
       eprint = {1609.07375},
 primaryClass = {astro-ph.HE},
       adsurl = {https://ui.adsabs.harvard.edu/abs/2017Sci...355..817I}
}

@ARTICLE{Becker2005,
       author = {{Becker}, Peter A. and {Wolff}, Michael T.},
        title = "{Spectral Formation in X-Ray Pulsar Accretion Columns}",
      journal = {\apjl},
         year = 2005,
        month = mar,
       volume = {621},
       number = {1},
        pages = {L45-L48},
          doi = {10.1086/428927},
archivePrefix = {arXiv},
       eprint = {astro-ph/0501434},
 primaryClass = {astro-ph},
       adsurl = {https://ui.adsabs.harvard.edu/abs/2005ApJ...621L..45B}
}

@ARTICLE{Paul2002,
       author = {{Paul}, B. and {Nagase}, F. and {Endo}, T. and {Dotani}, T. and {Yokogawa}, J. and {Nishiuchi}, M.},
        title = "{Nature of the Soft Spectral Component in the X-Ray Pulsars SMC X-1 and LMC X-4}",
      journal = {\apj},
         year = 2002,
        month = nov,
       volume = {579},
       number = {1},
        pages = {411-421},
          doi = {10.1086/342701},
archivePrefix = {arXiv},
       eprint = {astro-ph/0207341},
 primaryClass = {astro-ph},
       adsurl = {https://ui.adsabs.harvard.edu/abs/2002ApJ...579..411P}
}

@ARTICLE{Pradhan2021,
       author = {{Pradhan}, Pragati and {Paul}, Biswajit and {Bozzo}, Enrico and {Maitra}, Chandreyee and {Paul}, B.~C.},
        title = "{Comprehensive broad-band study of accreting neutron stars with Suzaku: Is there a bi-modality in the X-ray spectrum?}",
      journal = {\mnras},
         year = 2021,
        month = mar,
       volume = {502},
       number = {1},
        pages = {1163-1190},
          doi = {10.1093/mnras/stab024},
archivePrefix = {arXiv},
       eprint = {2101.01727},
 primaryClass = {astro-ph.HE},
       adsurl = {https://ui.adsabs.harvard.edu/abs/2021MNRAS.502.1163P}
}

@article{Esposito2016,
    author = {Esposito, P. and Israel, G. L. and Belfiore, A. and Novara, G. and Sidoli, L. and Rodríguez Castillo, G. A. and De Luca, A. and Tiengo, A. and Haberl, F. and Salvaterra, R. and Read, A. M. and Salvetti, D. and Sandrelli, S. and Marelli, M. and Wilms, J. and D'Agostino, D.},
    title = {EXTraS discovery of an 1.2-s X-ray pulsar in M 31},
    journal = {Monthly Notices of the Royal Astronomical Society: Letters},
    volume = {457},
    number = {1},
    pages = {L5-L9},
    year = {2016},
    month = {02},
    issn = {1745-3925},
    doi = {10.1093/mnrasl/slv194},
    url = {https://doi.org/10.1093/mnrasl/slv194},
    eprint = {https://academic.oup.com/mnrasl/article-pdf/457/1/L5/56942339/mnrasl_457_1_l5.pdf},
}

@article{Jain2024,
    author = {Jain, Chetana and Sharma, Rahul and Paul, Biswajit},
    title = {A comprehensive study of orbital evolution of LMC X-4: existence of a second derivative of the orbital period},
    journal = {Monthly Notices of the Royal Astronomical Society},
    volume = {529},
    number = {4},
    pages = {4056-4065},
    year = {2024},
    month = {04},
    issn = {0035-8711},
    doi = {10.1093/mnras/stae784},
    url = {https://doi.org/10.1093/mnras/stae784},
    eprint = {https://academic.oup.com/mnras/article-pdf/529/4/4056/57119981/stae784.pdf},
}

@ARTICLE{Tananbaum1972,
       author = {{Tananbaum}, H. and {Gursky}, H. and {Kellogg}, E.~M. and {Levinson}, R. and {Schreier}, E. and {Giacconi}, R.},
        title = "{Discovery of a Periodic Pulsating Binary X-Ray Source in Hercules from UHURU}",
      journal = {\apjl},
         year = 1972,
        month = jun,
       volume = {174},
        pages = {L143},
          doi = {10.1086/180968},
       adsurl = {https://ui.adsabs.harvard.edu/abs/1972ApJ...174L.143T}
}

@INPROCEEDINGS{Wilson1997,
       author = {{Scott}, D. Matthew and {Wilson}, Robert B. and {Finger}, Mark H. and {Leahy}, Denis A.},
        title = "{Observations of pulse evolution in Her X-1}",
    booktitle = {Proceedings of the Fourth Compton Symposium},
         year = 1997,
       editor = {{Dermer}, Charles D. and {Strickman}, Mark S. and {Kurfess}, James D.},
       series = {American Institute of Physics Conference Series},
       volume = {410},
        month = may,
    publisher = {AIP},
        pages = {748-752},
          doi = {10.1063/1.54005},
       adsurl = {https://ui.adsabs.harvard.edu/abs/1997AIPC..410..748S}
}

@ARTICLE{Corbet1984,
       author = {{Corbet}, R.~H.~D.},
        title = "{Be/neutron star binaries : a relationship between orbital period and neutron star spin period.}",
      journal = {\aap},
         year = 1984,
        month = dec,
       volume = {141},
        pages = {91-93},
       adsurl = {https://ui.adsabs.harvard.edu/abs/1984A&A...141...91C}
}

@ARTICLE{Corbet1986,
       author = {{Corbet}, R.~H.~D.},
        title = "{The three types of high-mass X-ray pulsator.}",
      journal = {\mnras},
         year = 1986,
        month = jun,
       volume = {220},
        pages = {1047-1056},
          doi = {10.1093/mnras/220.4.1047},
       adsurl = {https://ui.adsabs.harvard.edu/abs/1986MNRAS.220.1047C}
}

@ARTICLE{Burderi2010,
       author = {{Burderi}, L. and {Di Salvo}, T. and {Riggio}, A. and {Papitto}, A. and {Iaria}, R. and {D'A{\`\i}}, A. and {Menna}, M.~T.},
        title = "{New ephemeris of the ADC source 2A 1822-371: a stable orbital-period derivative over 30 years}",
      journal = {\aap},
         year = 2010,
        month = jun,
       volume = {515},
          eid = {A44},
        pages = {A44},
          doi = {10.1051/0004-6361/200912881},
archivePrefix = {arXiv},
       eprint = {1006.3283},
 primaryClass = {astro-ph.HE},
       adsurl = {https://ui.adsabs.harvard.edu/abs/2010A&A...515A..44B}
}

@ARTICLE{Jonker2001,
       author = {{Jonker}, Peter G. and {van der Klis}, Michiel},
        title = "{Discovery of an X-Ray Pulsar in the Low-Mass X-Ray Binary 2A 1822-371}",
      journal = {\apjl},
         year = 2001,
        month = may,
       volume = {553},
       number = {1},
        pages = {L43-L46},
          doi = {10.1086/320510},
archivePrefix = {arXiv},
       eprint = {astro-ph/0104356},
 primaryClass = {astro-ph},
       adsurl = {https://ui.adsabs.harvard.edu/abs/2001ApJ...553L..43J}
}

@ARTICLE{Heinke2015,
       author = {{Heinke}, C.~O. and {Bahramian}, A. and {Degenaar}, N. and {Wijnands}, R.},
        title = "{The nature of very faint X-ray binaries: hints from light curves}",
      journal = {\mnras},
         year = 2015,
        month = mar,
       volume = {447},
       number = {4},
        pages = {3034-3043},
          doi = {10.1093/mnras/stu2652},
archivePrefix = {arXiv},
       eprint = {1412.4097},
 primaryClass = {astro-ph.HE},
       adsurl = {https://ui.adsabs.harvard.edu/abs/2015MNRAS.447.3034H}
}

@ARTICLE{Degenaar2014,
       author = {{Degenaar}, N. and {Wijnands}, R. and {Reynolds}, M.~T. and {Miller}, J.~M. and {Altamirano}, D. and {Kennea}, J. and {Gehrels}, N. and {Haggard}, D. and {Ponti}, G.},
        title = "{The Peculiar Galactic Center Neutron Star X-Ray Binary XMM J174457-2850.3}",
      journal = {\apj},
         year = 2014,
        month = sep,
       volume = {792},
       number = {2},
          eid = {109},
        pages = {109},
          doi = {10.1088/0004-637X/792/2/109},
archivePrefix = {arXiv},
       eprint = {1406.4508},
 primaryClass = {astro-ph.HE},
       adsurl = {https://ui.adsabs.harvard.edu/abs/2014ApJ...792..109D}
}

@ARTICLE{hmxb_cat,
       author = {{Neumann}, M. and {Avakyan}, A. and {Doroshenko}, V. and {Santangelo}, A.},
        title = "{XRBcats: Galactic High Mass X-ray Binary Catalogue★}",
      journal = {\aap},
         year = 2023,
        month = sep,
       volume = {677},
          eid = {A134},
        pages = {A134},
          doi = {10.1051/0004-6361/202245728},
archivePrefix = {arXiv},
       eprint = {2303.16137},
 primaryClass = {astro-ph.HE},
       adsurl = {https://ui.adsabs.harvard.edu/abs/2023A&A...677A.134N}
}

@ARTICLE{lmxb_cat,
       author = {{Avakyan}, A. and {Neumann}, M. and {Zainab}, A. and {Doroshenko}, V. and {Wilms}, J. and {Santangelo}, A.},
        title = "{XRBcats: Galactic low-mass X-ray binary catalogue}",
      journal = {\aap},
         year = 2023,
        month = jul,
       volume = {675},
          eid = {A199},
        pages = {A199},
          doi = {10.1051/0004-6361/202346522},
archivePrefix = {arXiv},
       eprint = {2303.16168},
 primaryClass = {astro-ph.HE},
       adsurl = {https://ui.adsabs.harvard.edu/abs/2023A&A...675A.199A}
}
\bibliographystyle{aasjournalv7}


\end{document}